\documentclass[12pt,a4paper,final]{iopart}

\usepackage{iopams}  
\usepackage{graphicx}
\usepackage[breaklinks=true,colorlinks=true,linkcolor=blue,urlcolor=blue,citecolor=blue]{hyperref}
\usepackage{color}
\usepackage{subfigure}

\begin{document}

\title{Spin transport in dangling-bond wires on doped H-passivated Si(100)}

\author{Mika\"el Kepenekian$^{1,2}$, Roberto Robles$^{2,3}$,
	Riccardo Rurali$^4$, Nicol{\'a}s Lorente$^{2,3}$}

\address{$^1$Institut des Sciences Chimiques de Rennes UMR 6226, CNRS -
	Universit{\'e} de Rennes 1, Rennes, France}
\address{$^2$ICN2 - Institut Catala de Nanociencia i Nanotecnologia, Campus
	UAB, 08193 Bellaterra (Barcelona), Spain}

\address{$^3$CSIC - Consejo Superior de Investigaciones Cientificas, ICN2
	Building, Campus UAB ,08193 Bellaterra (Barcelona), Spain}

\address{$^4$Institut de Ci{\`e}ncia de Materials de Barcelona (ICMAB-CSIC),
	Campus de Bellaterra, 08193 Bellaterra (Barcelona), Spain}


\ead{mikael.kepenekian@univ-rennes1.fr}

\begin{abstract}
	New advances in single-atom manipulation are leading to the creation
	of atomic structures on H passivated Si surfaces with functionalities
	important for the development of atomic and molecular based
	technologies. We perform total-energy and electron-transport calculations
	to reveal the properties and understand the features of atomic wires
	crafted by H removal from the surface. The presence of dopants radically
	change the wire properties. Our calculations show that dopants have a
	tendency to approach the dangling-bond wires, and in these conditions,
	transport is enhanced and spin selective. These results have important
	implications in the development of atomic-scale spintronics showing that
	boron, and to a lesser extent phosphorous, convert the wires in high-quality
	spin filters.
\end{abstract}


\vspace{2pc}
\submitto{\NT}
\maketitle

\section{Introduction}
The integration of semiconductors and magnetic materials is of great
importance for the creation of new technology in which digital data are
encoded in the spin of electrons.~\cite{jansen2012a}
To this respect, the use of molecules that can become spin devices is
a very interesting possibility.~\cite{sanvito2011a} Indeed, molecular
devices are very small, adaptable and easy to create using chemical
engineering. Their spin functionalities are also very interesting, and
they have been deemed superior to more common spintronic
strategies.~\cite{rocha2005a,bogani2008a} As in molecular
electronics,~\cite{aviram1974a, joachim2000a} the problem still
comes from creating efficient atomic-size interconnects to form
complete circuits.~\cite{joachim2000a,kepenekian2013b}

Interconnects that can transport spin are then desirable for spintronic
application. Promising candidates for surface interconnects can be
found in carbon nanotubes (CNTs),~\cite{iijima1991a, charlier2007a}
as they present both easy synthesis and complex physics. In particular,
the low Z of carbon, assures very small spin-orbit coupling and hence
long spin lifetimes. However, an important issue with CNTs is that their
properties are extremely sensitive to the topological structure of the tube.
Other serious candidates emerge from the variety of silicon nanowires
(Si NWs) that can be grown and that happen to present an easier control
on their physical properties.~\cite{cui2001a, rurali2010a}
A new type of nanowires has been recently proposed with embedded
phosphorous in a silicon crystal. The resulting 1-D system exhibits then
a very low resistivity.~\cite{weber2012a}
An alternative path consists in using the scanning tunneling microscope
(STM) to selectively remove hydrogen atoms from a H-passivated Si(100)
surface along the Si dimer row leading to a dangling-bond (DB)
wire.~\cite{shen1995a, hosaka1995a, hitosugi1999a, soukiassian2003a,
hallam2007a, haider2009a, pitters2011a}
At variance with isolated DBs, which introduce localized mid-gap states,
these wires give rise to dispersive bands with a marked one-dimensional
character.
The stability and the transport properties of such wires have been extensively
studied by tight-binding as well as ab intitio methods~\cite{watanabe1996a,
watanabe1997a, doumergue1999a, cho2002a, bird2003a, cakmak2003a,
lee2008a, lee2009a, kawai2010a, lee2011a, robles2012b, kepenekian2013a}
and several experimental proofs-of-concept have been
reported.~\cite{hitosugi1999a, soukiassian2003a, schofield2013a}. Spin
lifetimes in silicon are known to be long, however DB wires present some
problems for transport: $(i)$ transport properties of simple 1-D wires are
limited due to the appearance of an electronic
gap,~\cite{kepenekian2013a, kepenekian2013b} and $(ii)$ the coupling of
the wire with the substrate due to the presence of dopants and impurities
leads to a leak of the electronic current in the bulk.~\cite{kepenekian2013c}

In the present work we focus on the effect of dopants, boron ($p$-type) and
phosphorous ($n$-type), on the electronic transport properties of DB-wires
with special interest in the spin transport properties.
In a semiconductor, the density of carriers can be tuned by adding dopants. 
In a conventional approach, DBs are undesired surface defects, because
they introduce deep states in the band-gap that act as charge traps, reducing
the conductivity. 
Here, we place ourselves in the opposite situation: DB conductive wires are
engineered  on a Si surface, for a specific purpose, e.g. transporting charge
and spin between two devices, and we focus on how dopants affect transport
in these wires.
We start by analyzing the dopant segregation on the H-passivated Si(100)
surface with and without DB wires, finding that dopants segregate at the
surface.
Next, we observe two remarkable effects of the dopants: first, dopants greatly
reduce the gap, which leads to  quasi-metallic wires; second, as they stabilize
the magnetic ordering in the wires,  the wire transport becomes spin polarized.
We find that DB-wires are good spin filters, going beyond simple interconnects.
Hence, DB wires will not only transport spin, but will make sure that the spin
polarisation is preserved.

\section{Computational details}
First-principles calculations are based on density functional theory (DFT) as
implemented in the {\sc Siesta} package.~\cite{soler2002a, artacho2008a}
Calculations have been carried out with the GGA functional in the PBE
form,~\cite{perdew1996a}
Troullier-Martins pseudopotentials,~\cite{troullier1991a} and a basis set of
finite-range numerical pseudoatomic orbitals for the valence wave
functions.~\cite{artacho1999a}
Structures have been relaxed using a double-$\zeta$ polarized basis
sets.~\cite{artacho1999a}
The surfaces were modeled using a slab geometry with eight silicon layers and
a $4 \times 8$ unit cell of the H-passivated Si(100)-(2$\times$1) surface.
The extent of the cell allows one to limit the direct interaction between dopants,
but still leads to a very high concentration of dopant with one Si atom out of 600
being substituted by a B or P atom.
The electronic structure was converged using a $1 \times 5 \times 5$ $k$-point
sampling of the Brillouin zone.
Conductances have been computed using a single-$\zeta$ polarized basis set,
by means of the {\sc TranSiesta} method,~\cite{brandbyge2002a} within the
non-equilibrium Green's function (NEGF) formalism.
The following setup has been used: 4-dimer DB wires act as left and right
electrodes, while a 8-dimer DB wire constitutes the scattering region.
The current is evaluated following Landauer's equation:~\cite{datta_book}
\begin{equation*}
	I = \frac{2 e}{h} \int_{-\infty}^{\infty} T(E,V) [ f_R(E)-f_L(E)] dE
\end{equation*}
where $T(E,V)$ is the transmission function for an electron of energy $E$
when the bias between the two DB electrodes is $V$, and $f_R(E)$ ($f_L(E)$)
is the right- (left-) electrode Fermi occupation function. We further simplify
the current $I$ calculation using the zero-bias transmissions.
In {\sc TranSiesta}, bias between the electrodes is established by applying an
electric field along the transport direction. Mobile charges inside the electrodes
screen the electric field, resulting in the usual electric potential profile (flat inside
the electrodes, potential drop in the scattering region).
Only electrodes that are three-dimensional metals guarantee this kind of screening.
In this configuration, however, it becomes very difficult to disentangle the
contributions to the resistance of the electrode-semiconductor interface from those
of the sub-surface dopant scattering.
We believe that our simplified model gives a better insight into the physics of
electron transport through DB wires in presence of dopant, because all
the observed effects (stabilization of a magnetic solution, leakage, \dots) necessarily
stem from the presence of the impurities. The price to pay is that biased calculation
cannot be performed for the methodological reasons outlined above.
In all cases, an energy cutoff of 200 Ry for real-space mesh size has been used.

\section{Results and discussion}

\begin{figure}[t]
	\begin{center}
		\includegraphics*[width=.5\linewidth]{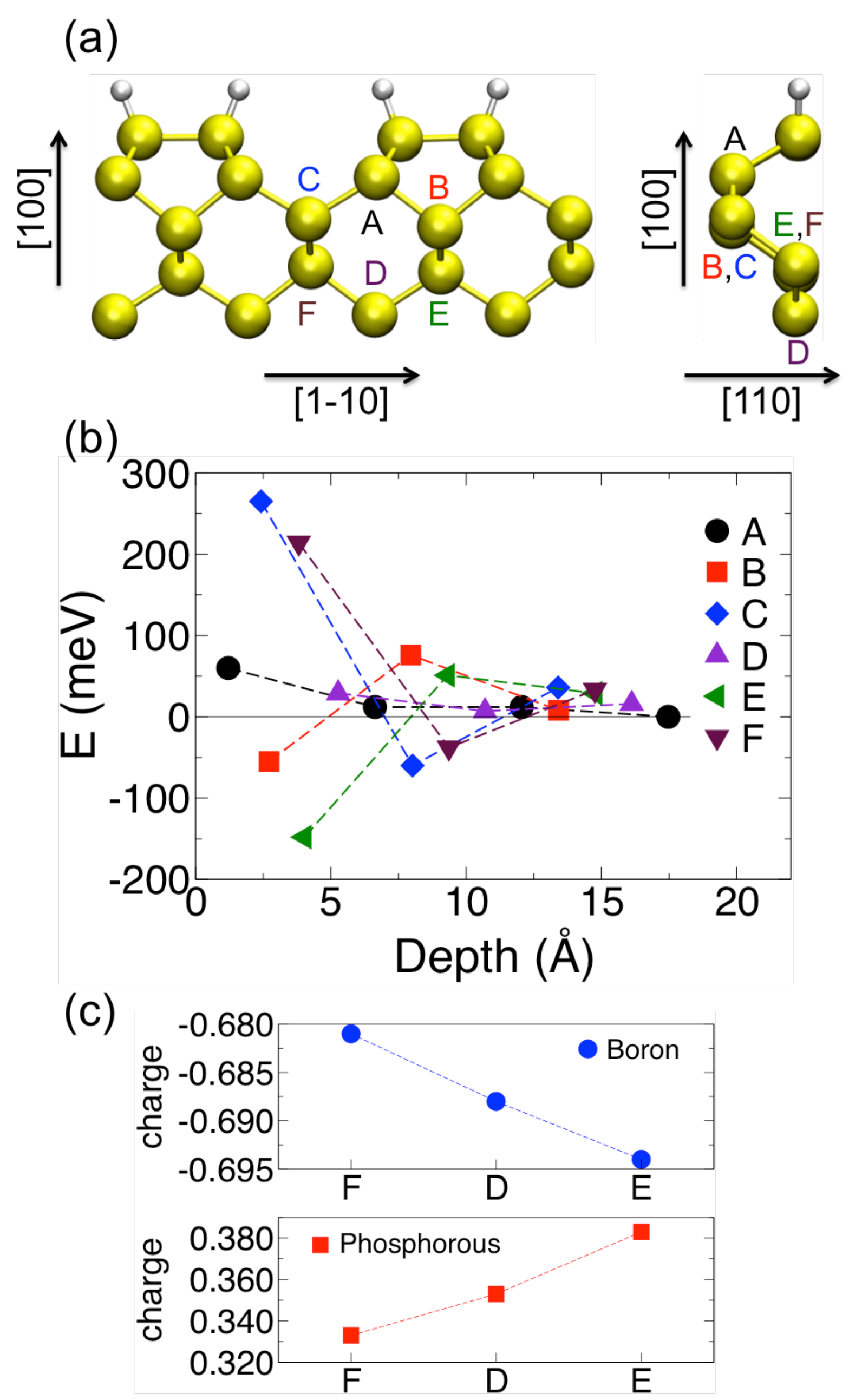}
	\end{center}
	\vspace{-0.3cm}
	\caption{(a) Side views of the H-passivated Si(100) surface, where the
		dimers formed on the surface can be seen on the upper part. Si
		and H atoms are depicted in yellow and white, respectively.
		The different lattice sites
		for B and P are indicated. Sites A, B and C are shifted with respect to
		the surface dimers, whereas sites D, E and F are positioned below
		them (cf. right side).
		(b) Formation energies (meV) of boron in Si close to the H-passivated
		Si(100) surface. At the bulk limit all the sites A, B, C, D, E, and F are
		equivalent. The corresponding formation energy is taken as a reference.
		The formation energies become site-specific close to the surface.
		(c) Calculated Mulliken net charges of the boron and phosphorous
		impurities depending on the site substituted. Dashed lines are a
		guide to the eye.}
	\label{fig:segregation}
\end{figure}

The impact of dopants on the current carried by DB-wires depends critically on
their distribution with respect to the surface. Hence, our first goal was studying
the surface segregation of these defects.
Broadly speaking, we find that in this system impurities prefer to
be closer to the surface because the strain introduced can be more
easily released. This is a well-known behaviour in similar systems, like
unpassivated Si surfaces~\cite{luo2003a, centoni2005a} and in Si
nanowires.~\cite{fernandez-serra2006a,peelaers2006a,rurali2010a}
Our calculations show the same trend in the H-passivated Si(100) surface.
For both B and P, nearly 150 meV are gained in the most stable surface
position (see Figure~\ref{fig:segregation}(b) and Table~\ref{tab:segregation}).
As the dopant gets closer to the surface, the formation energy becomes much
more site-specific.
It appears that sites D, E and F are favored with respect to
sites A, B and C, respectively.
These differences vanish quickly moving away
from the surface and are within the numerical accuracy of the calculation at
15~\AA\ from the surface. In the following, we limit our discussion to the case
of a substitution at sites D, E and F which appears to be more stable than
those at A, B and C sites.

\begin{table}[htbp]
	\begin{center}
		\caption{\label{tab:segregation} Formation energies (in meV, see text)
			for sites D, E, and F with and without DB-wires. The notations
			X@H-passivated, X@NM and X@AFM stands for a system with
			the dopant X (X=B, P) below a H-passivated surface, a NM and
			AFM wire, respectively.}
		\begin{tabular}{lrrrrrr}
			\hline
			\hline
	 && D && E && F \\
			\hline
	B@H-passivated	&& 29 		&& $-148$	&& 214		\\
	B@NM			&& $-14$		&& $-415$	&& $-12$		\\
	B@AFM			&& $-10$		&& $-110$	&& $-123$	\\
	\hline
	P@H-passivated	&& $-61$		&& $-145$	&& $105$		\\
	P@NM			&& $-237$	&& $-488$	&& $-57$		\\
	P@AFM			&& $-329$	&& $-539$	&& $-180$	\\
			\hline
			\hline
		\end{tabular}
	\end{center}
\end{table}

The site selectivity observed in Figure~\ref{fig:segregation}(b) depends on
several intertwined factors (strain relaxation, electronic states of the
neighbouring DBs...),  but can be roughly tracked back
to the amount of charge transfer by the dopant depending on the site
substituted.
We find that the most stable site corresponds to the largest charge
transfer, {\it i.e.} to the strongest bond form between the dopant and its Si
neighbours (see Figure~\ref{fig:segregation}(c)).

Starting from the previous structures, DB-wires are {\it drawn} on the surface
by removing a row of Hydrogens along the [110] direction.
In the absence of dopants, such wires are known to be unstable and relax
following either a non-magnetic (NM) Peierls distortion or a spin-polarized
solution that leads to  antiferromagnetic (AFM) ordering.~\cite{watanabe1996a,
watanabe1997a, cho2002a, bird2003a, cakmak2003a, lee2008a, lee2009a,
lee2011a, robles2012b}
The two solutions are very close in energy with the NM configuration more
stable than the AFM one by 5 meV/DB.~\cite{robles2012b}

Our results show that surface segregation is enhanced by the presence of
such DB-wires. Indeed, whatever the site of substitution for B or P is, the
formation energy is smaller than in the bulk position (see
Table~\ref{tab:segregation}). This is in agreement with previous results on Si
NWs,~\cite{fernandez-serra2006a, fernandez-serra2006b} where 
dopants are found to form electrically inactive complexes with isolated DBs.
Also here, the tendency to surface segregation is more pronounced with P,
as observed in the case of Si NWs and unpassivated Si(100)
surface.~\cite{fernandez-serra2006a, fernandez-serra2006b}

\begin{table}[hbt]
	\begin{center}
		\caption{\label{tab:afmnm} Energy differences (in meV/DB) between
			NM and AFM configurations for top sites D, E, and F. In the
			absence of dopant this difference is $-5$ meV/DB.~\cite{robles2012b}}
		\begin{tabular}{cccccccccccc}
			\hline
			\hline
	B@D && B@E && B@F &&& P@D && P@E && P@F\\
			\hline
		10 && 22 && 16 &&&
		20 && 15 && 21 \\
			\hline
			\hline
		\end{tabular}
	\end{center}
\end{table}

More interestingly, we have found that both B and P stabilize the magnetic
solution with respect to the NM distorted one, regardless of the site of
substitution (see Table~\ref{tab:afmnm}).
The destabilization of the NM wire is due to the increased distortion caused
by the dopant. The shortening going from Si-Si to Si-B or Si-P bonds is
conflicting with the buckling imposed by the NM Peierls-like structure. As an
example the buckling in the non-doped NM structure gives differences of
$\Delta z =$0.69~\AA\ in the vertical direction between neighbour DBs. In a
B-doped (P-doped) system, this height difference becomes as high as
$\Delta z =$1.34~\AA\ ($\Delta z =$1.52~\AA) close to the dopant.
In the undistorted AFM wire the deformation introduced by the dopant is
much less: $\Delta z$ of 0.16~\AA\ and 0.24~\AA\ for B and P, respectively.
The geometry of the magnetic solution is preserved. Thus, in the presence
of dopants the AFM solution becomes the ground state, yielding a magnetic
ordering that can be exploited for spin transport related applications.

Dopants lead to an injection of hole or electrons. In the presence of DBs on
H-passivated Si(100), the extra charges are trapped by these surface defects,
leading to a decrease in the conductance. Here, however, where conduction
is supposed to take place along the DB-wires, this effect turns out to be
positive by closing the electronic gap of the AFM wire and leading to a
spin-specific quasi-metallicity.

\begin{figure}[t]
	\begin{center}
		\includegraphics*[width=.65\linewidth]{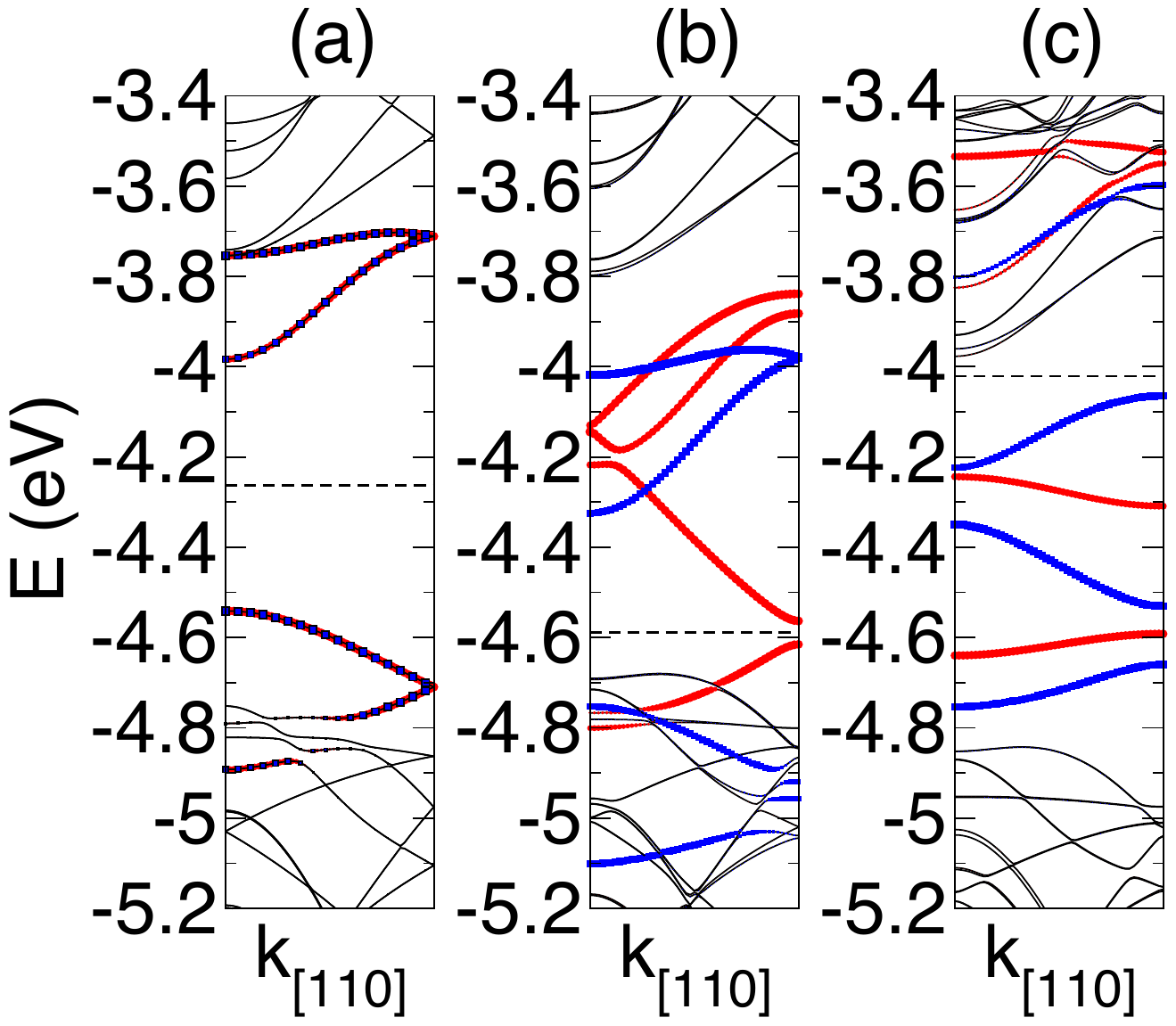}
	\end{center}
	\vspace{-0.3cm}
	\caption{Electronic bands evaluated for a periodic magnetic nanowire with
		(a) a pure Si substrate, (b) B-doped Si substrate, and (c) P-doped Si
		substrate. The red and blue circles indicate the weight of the DB in
		the majority and minority spin bands, respectively.
		The dashed lines indicate the position of the Fermi energy.
		}
	\label{fig:bands}
\end{figure}

Figure~\ref{fig:bands} shows the band structure of (a) an undoped AFM wire,
(b) an AFM wire with a substitutional B atom in its most stable configuration
(F site) and (c)  an AFM wire with a substitutional P atom in its most stable
configuration (E site). One can see the expected shifting of the Fermi energy
(dashed line) controlled by the amount of extra charge/hole injected in the
system through the dopant.  The undoped AFM wire, Fig.~\ref{fig:bands} (a),
shows two surface states leaving a surface-band gap of 0.56~eV. For each
surfaces state, the bands corresponding to each spin overlap due to the AFM
ordering.
The B-doped system, Fig.~\ref{fig:bands} (b), displays a splitting of bands
according to spin  (red and blue for majority and minority spins). The splitting
of bands  is due to the introduction of an extra spin in an otherwise perfect
AFM wire, leading to an unbalanced number of spins. The increase of the DB
charge, also caused by the dopant, produces the reduction of the surface-band
gap to 0.05 eV. In the case of P doping, the extra spin also produces the spin
polarisation of bands and a 0.09-eV gap. Therefore, the presence of dopants
brings the initial insulating system to a spin-polarized quasi-metallic state.

\begin{figure}[t]
	\begin{center}
		\includegraphics*[width=.55\linewidth]{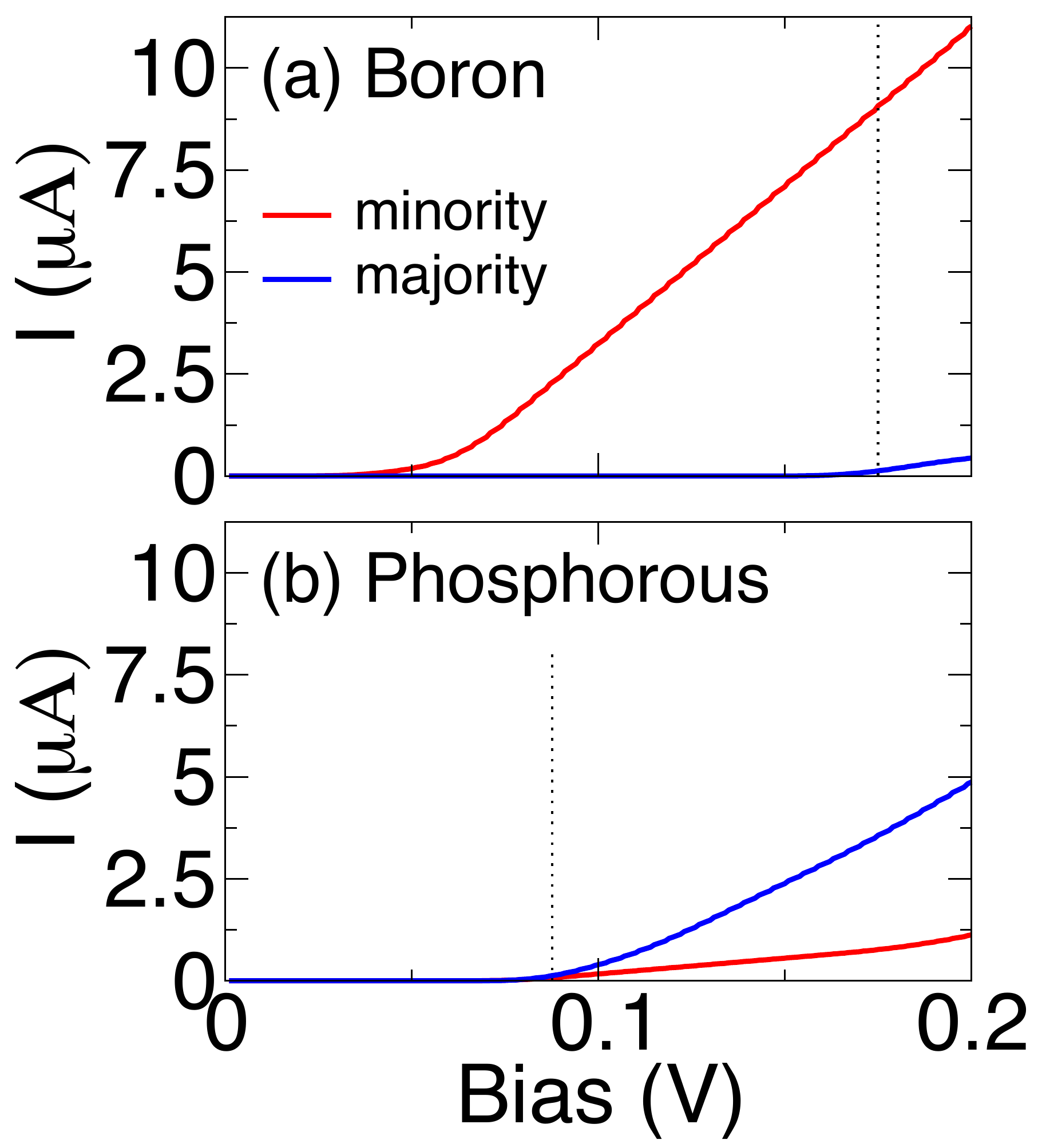}
	\end{center}
	\vspace{-0.3cm}
	\caption{I-V curve for the most stable B-doped (a) and P-doped (b) AFM
		dangling-bond wire. The doted black line indicates the bias at which
		bulk contributions to the current starts. For larger biases, a part of
		the current is lost into the Si bulk (leakage current). The non-doped
		AFM DB wire presents a gap larger than the bias window considered
		here.}
	\label{fig:current}
\end{figure}

Figure~\ref{fig:current} shows the computed I-V curves for B- and P-doped AFM
wires. In both cases, the bias required to obtain a current response is 
below 0.1 V in agreement with the above electronic gaps.
In the case of P doping, the current contains a bulk contribution to the current.
This leads to leakage current, {\it i.e.} a loss of surface current into the Si
bulk. A previous study showed that this loss could represent as much as 30\%
of the total current for low biases (less than 0.5 V).~\cite{kepenekian2013c}
Figure~\ref{fig:current} displays the current along the DB wire that does
not contain the fraction of the current lost into the bulk. This explains why
the current is largely spin polarized despite having a larger number of bulk
bands, because the wire current is mainly due to the first band
below the Fermi energy, which is a surface band and hence
spin polarized.

However, current leakage is negligible for B doping because there is no bulk
bands in the energy windows for low biases. Indeed, for biases between 0.05
and 0.18 V the current remains on the surface and the current leakage is
strictly zero.
The spin-polarisation for both dopings is the same,  {\it i.e.} adding or subtracting
one electron by the dopant will change the spin balance on the DB, but the majority
spin will be the same spin. However, the surface current shows different
spin-polarisations, Fig.~\ref{fig:current}. This is due to the actual ordering of the
DB bands, which changes under the dopant potential.

\begin{figure}[t!]
	\begin{center}
		\includegraphics*[width=.5\linewidth]{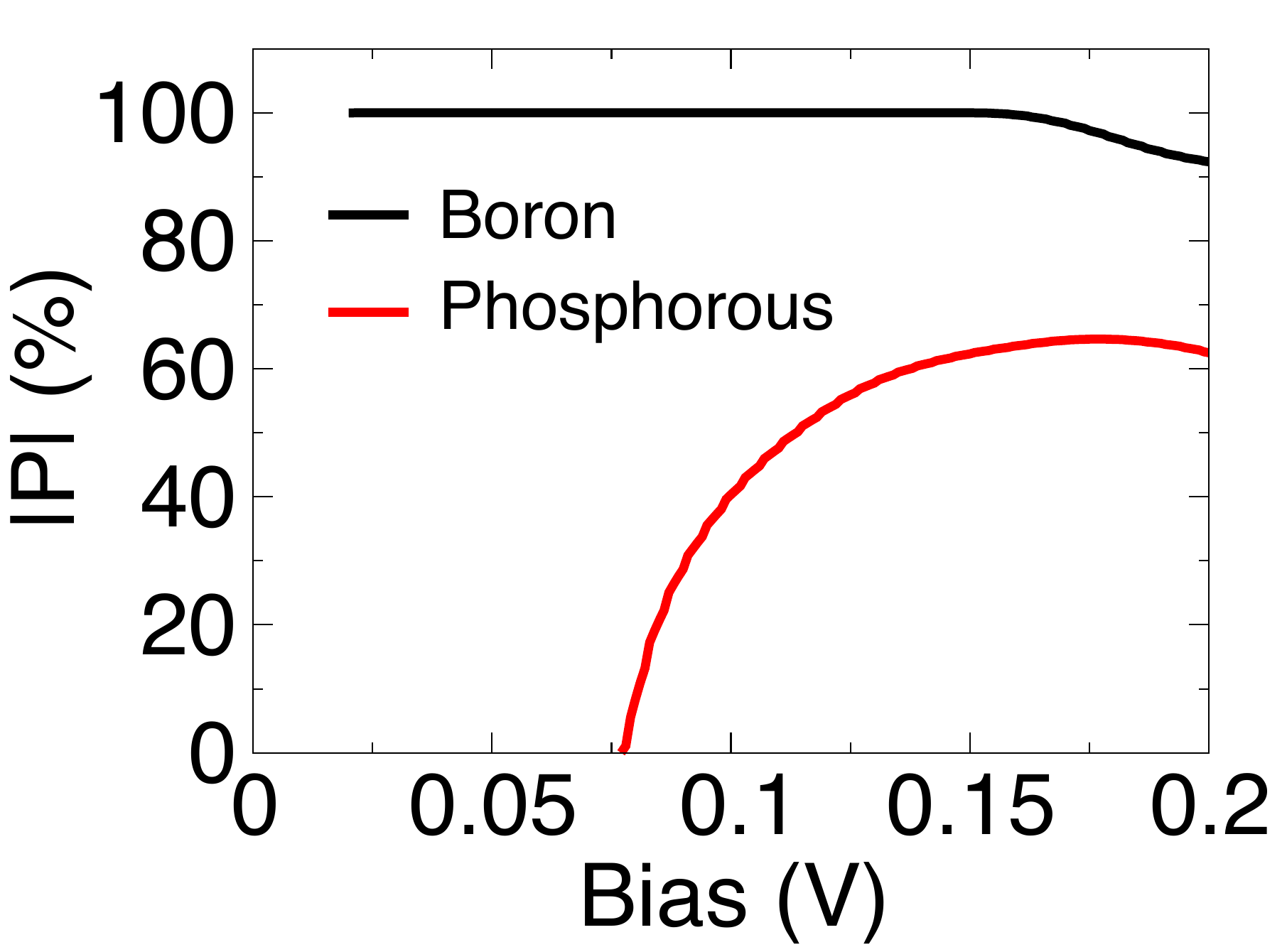}
	\end{center}
	\vspace{-0.3cm}
	\caption{Spin polarisation for the B-doped system (black) and the
		P-doped one (red). The curves are not defined for biases below the
		band gaps, since the wires hold no current. B-doped systems show
		100\% polarisation for biases below 0.17 V. Beyond this bias the
		presence of bulk bands contribute to current leakage and to the loss
		of spin polarisation. In the case of P-doped system, the  bulk bands
		contribution starts as early as 0.09 eV, leading to a lower spin
		polarisation. Therefore, for biases lower than 0.17 V, the B-doped
		DB wire drawn on H-passivated Si(100) system is a perfect
		non-leaking spin-filtering surface interconnect.}
	\label{fig:polarisation}
\end{figure}

By defining the spin polarisation like $P= (I_\uparrow - I_\downarrow)/(I_\uparrow + I_\downarrow)$,
we obtain that B-doped systems present a 100\% polarisation for biases below
0.17 V, Fig.~\ref{fig:polarisation}.
Beyond this bias the presence of bulk bands contribute to current leakage and
to the loss of spin polarisation. In the case of P-doped system, the  bulk bands
contribution starts as early as 0.09 eV, leading to a lower spin polarisation.
Therefore, for biases lower than 0.17 V, the B-doped DB wire drawn on
H-passivated Si(100) system is a perfect spin-filtering surface interconnect
thanks to the absence of current leakage and to the perfect spin polarisation.

\section{Conclusion}
In summary, we have shown that boron and phosphorous dopants segregates
to the H-passivated Si(100) surface. This phenomenon is enhanced by the
presence of of DB wires. The first effect of dopants is to stabilize the magnetic
form of DB wires over the non-magnetic Peierls distorted one.
As observed in other doped Si systems, the extra charge brought by
the dopant is captured by the DB wire. One consequence is the closing of the
electronic gap leading to quasi-metallicity of AFM DB wires.
Moreover, the presence of dopants induce a total magnetic moment on the
electronic bands close to the Fermi energy, leading to spin-specific electron
transport.
In the case of B-doped AFM DB wire, the current is not only free of leakage
from the wire but also spin-specific. Therefore, B-doped DB wires are perfect
spin-filtering surface interconnects.

\section*{Acknowledgments}

This work has been supported by the European Union Integrated Project AtMol
(http://www.atmol.eu).
R. Rurali acknowledges funding under contract Nos. FEDER-FIS2012-37549-C05-05
and CSD2007-00041 of the Ministerio de Econom\'ia y Competitividad (MINECO).
R. Robles and N. Lorente acknowledge financial support from Spanish MINECO
(Grant No. MAT2012-38318-C03-02 with joint nancing by FEDER Funds from the
European Union).


\section*{References}


\end{document}